\documentclass[twocolumn,showpacs,preprintnumbers,amsmath,amssymb]{revtex4}

\usepackage{graphicx}
\usepackage{dcolumn}
\usepackage{bm}

\def\lapproxeq{\lower .7ex\hbox{$\;\stackrel{\textstyle
<}{\sim}\;$}}
\def\gapproxeq{\lower .7ex\hbox{$\;\stackrel{\textstyle
>}{\sim}\;$}}

\begin{document}

\title{
Upper limit on the photon fraction in highest-energy cosmic rays from AGASA data
}

\author{
M.~Risse$^{1,2,*}$}
\author{
P.~Homola$^1$}
\author{
R.~Engel$^2$}
\author{
D.~G\'ora$^1$}
\author{
D.~Heck$^2$}
\author{
J.~P\c{e}kala$^1$}
\author{
B.~Wilczy\'nska$^1$}
\author{
H.~Wilczy\'nski$^1$}
\affiliation{
(1) Institute of Nuclear Physics, Polish Academy of Sciences, 
ul.~Radzikowskiego 152, 31-342 Krak\'ow, Poland \\
(2) Forschungszentrum Karlsruhe, Institut f\"ur Kernphysik,
76021~Karlsruhe, Germany \\
*Corresponding author. Electronic address: markus.risse@ik.fzk.de
}


\begin{abstract}

A new method to derive an upper limit on photon primaries from small
data sets of air showers is developed which accounts for shower 
properties varying with the primary energy and arrival direction. 
Applying this method to the highest-energy showers recorded by the 
AGASA experiment,
an upper limit on the photon fraction of 51\% (67\%) at a confidence
level of 90\% (95\%) for primary energies 
above $1.25\cdot 10^{20}$~eV is set.
This new limit on the photon fraction above the GZK cutoff energy
constrains the Z-burst model of the origin of highest-energy cosmic rays.

\end{abstract}

\pacs{96.40.Pq,96.40.-z,13.85.-t,13.85.Tp}

\keywords{Suggested keywords}

\maketitle

Since their first discovery about 40 years ago~\cite{vulcano},
the existence of particles with energies around and above 
100~EeV = $10^{20}$~eV
was confirmed by several air shower experiments using different measurement
techniques~\cite{flyseye,agasa-213,havpark,
agasa-data,hires,yakutsk}.
The quest for the nature and origin of these extremely high-energy (EHE)
cosmic rays continues to drive considerable experimental
and theoretical efforts 
\cite{reviews}.
As first pointed out by Greisen, Zatsepin and Kuzmin~\cite{gzk},
the travel distance
of EHE particles is limited due to energy losses on background radiation
fields. 
For instance, the energy loss length of 200~EeV protons is
$\simeq$30~Mpc (e.g.~\cite{reviews,stecker}).
A cosmological origin of the observed EHE particles thus seems disfavoured.
However, no astrophysical object in our cosmological vicinity could
be identified yet as source of these events.
Moreover, explaining efficient particle acceleration to such enormous
energies poses a theoretical challenge.
The acceleration problem is circumvented if EHE particles are generated in
decays or annihilation of topological defects (TD)
or super-heavy dark matter 
(SHDM)~\cite{exotic,berez04,semsig,bhat-sigl}. 
These objects are expected in certain inflation scenarios and have
also been proposed as dark matter candidates.

A common feature of such non-acceleration models 
is the large fraction of
EHE photons in the injected particle spectrum.
Due to interactions of these photons with background 
fields, the diffuse photon flux at GeV energies
allows one to derive an upper limit on the
electromagnetic energy injected as EHE particles at distances beyond a
few Mpc~\cite{constr_princ,egret-diff}.
This constrains non-acceleration models
which predict
particle injection at large
distances~\cite{sem-sigl,berez04}.
In turn, 
models with injection sites closer to the
observer
imply a significant fraction of primary photons in the observed EHE
events.
As an example,
in the SHDM model metastable particles of mass
$M_x \simeq 10^{14}$~GeV are clumped in the galactic halo and produce
EHE photons, nucleons and neutrinos by decay~\cite{berez04}.
Thus, 
stringent limits on the EHE photon fraction 
provide constraints on non-acceleration models complementary
to those
from the
GeV photon background.

Based on an analysis of muons in air showers observed by the
Akeno Giant Air Shower Array (AGASA), upper limits on the photon fraction
were estimated to be 28\% above 10~EeV and 67\% above 32~EeV
(95\% CL)~\cite{agasa-data}.
Comparing rates of near-vertical showers to inclined ones,
upper limits of
48\% above 10~EeV and 50\% above 40~EeV
(95\% CL) were deduced
from Haverah Park data~\cite{havpark}.
Non-acceleration models of cosmic-ray origin are not severely
constrained by these 
bounds~\cite{kachelr}, however.
Photons are predicted to reach a considerable fraction
only at highest energies, while with decreasing 
energies below 100~EeV the ``conventional'' hadronic cosmic-ray component
soon outnumbers photon primaries due to the steep flux spectrum.
For instance, based on the SHDM model the photon fraction above 40~EeV
is estimated as $\simeq$25\% only, increasing to $\simeq$50\% at 
70~EeV~\cite{berez04}.

In this work, we focus on events above 100~EeV. These particles
are most directly linked to the production scenario in
non-acceleration models, and the predicted photon dominance can be
checked with the data.
The largest data set on EHE events available to date was obtained by
the AGASA experiment.
From eleven AGASA showers reconstructed with energies above
100~EeV, the muon content
in the shower is measured in six~\cite{agasa-excess,agasa-data}.
For each event, adopting its reconstructed primary parameters,
we compare the observed muon signal to results from
shower simulations for primary photons.
In contrast to the analysis method in~\cite{agasa-data},
where the data distribution above energies of 10~EeV and 32~EeV
was compared to an overall simulated
distribution, we thus
use the information about the individual event characteristics.
We develop a new statistical method that allows us to combine
the information from all events and to set a limit on the
primary photon contribution.

AGASA~\cite{agasa-data,agasa}
consisted of 111 array detectors spread over $\simeq$100~km$^2$ area
and 27 muon detectors with an energy threshold of 0.5~GeV for vertically
incident muons.
The primary energy was determined 
with a statistical accuracy of $\simeq$25\% for
hadron primaries~\cite{agasa-excess}.
Assuming photon primaries, the energies reconstructed this way were found to
be underestimated by $\simeq$20\% for the most-energetic 
events~\cite{agasa-data}.
Six events were reconstructed with $>$100~EeV which had
more than one muon detector within 800-1600~m distance from
the shower core~\cite{agasa-data}. The muon density
$\rho_j$
at 1000~m core
distance was obtained for each event $j$=$1\dots 6$
by fitting an empirical lateral
distribution function~\cite{latdistr} to the data.
The uncertainty estimated for the resulting $\rho_j$ is 40\% \cite{agasa-data}.
The reconstructed 
shower parameters of the highest-energy events with muon data
are given in Tab.~\ref{table1}.
\begin{table}[t]
\begin{center}
\caption{Reconstructed shower parameters of the AGASA 
events~\cite{agasa-data} (upper part of the Table) 
and simulation results (lower part). 
The energies are increased by 20\% to account for the case of photon
primaries~\cite{agasa-data}.
The azimuth angle is given clockwise from north for the incoming
direction.} 
\label{table1}
\begin{tabular}{ccccccc}
\hline
primary energy [EeV]      &295   & 240    &  173   &  161   &  126   &  125
\\
zenith angle [$^{\circ}$] & 37 & 23   & 14   & 35   & 33   & 37
\\
azimuth angle [$^{\circ}$]& 260  & 236    &  211   &   55   &  108   & 279
\\
$\rho_j$ [m$^{-2}$]&8.9  &  10.7  & 8.7    & 5.9    & 12.6   & 9.3
\\
\hline
preshower occurrence [\%]    &100  & 100   &   96   & 100    &   93   & 100
\\
$<$$\rho_j^{\rm s}$$>$ [m$^{-2}$]&4.2&3.1& 2.1 & 2.3  & 1.7  & 1.8
\\
$\Delta \rho_j^{\rm s}$ [m$^{-2}$]&1.1&1.0&0.9 & 0.6  & 0.5   & 0.5
\\
$\chi_j^2$      &  1.6  & 3.0 &   3.4  &  2.2   &  4.6   & 4.0
\\
$p_j$ [\%]      &  20.8 & 8.3 &   6.4  &  13.9  &  3.1   & 4.6
\\
\hline
\end{tabular}
\vskip -0.8 cm
\end{center}
\end{table}

\begin{figure}
\centerline{
\includegraphics[width=7.3cm]{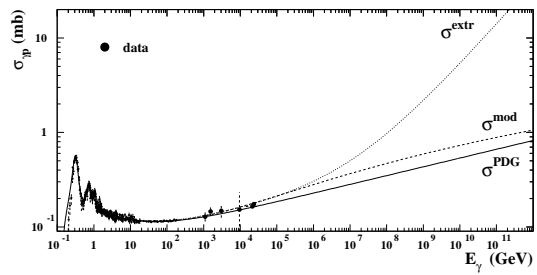}
}
\vskip -0.3 cm
\caption{Data~\cite{pdg} and
extrapolations of the photonuclear cross-section 
$\sigma_{\gamma p}$.
The PDG extrapolation ($\sigma^{\rm PDG}$)~\cite{pdg,cud}
is chosen for this analysis.
Also shown are two parametrizations with larger
cross-sections at highest energies, 
denoted $\sigma^{\rm mod}$~\cite{bezbug} and
$\sigma^{\rm extr}$~\cite{donland} (see text).
The cross-section on air is taken as
$\sigma_{\gamma - \rm air}$ = 11.44 
$\sigma_{\gamma p}$~\cite{corsika,landboern}.}
\label{fig-sigma_phn}
\vskip -0.4 cm
\end{figure}

It is well known that
photon-initiated showers generally contain
significantly fewer muons than hadron-induced events.
For each AGASA event, 100 primary photon showers were generated.
The reconstructed primary direction~\cite{agasa-excess}
was chosen as simulation input and the primary energy varied from
shower to shower according to the reconstruction uncertainty.
The energy was also globally increased by 20\% to account for the
energy underestimation in case of photons.
Electromagnetic cascading of photons in the geo\-magnetic field
was simulated for the AGASA site with the new
PRESHOWER code~\cite{preshower}.
For the 
AGASA events, in most cases
preshower formation occurred (see Tab.~\ref{table1}).
The resulting atmospheric shower was simulated with 
CORSIKA~6.18~\cite{corsika} as a superposition
of subshowers initiated by the preshower particles or, if no preshower
occurred, with the original primary photon.
Electromagnetic interactions were treated by the EGS4 code~\cite{egs},
which was upgraded~\cite{corsika} to take photonuclear reactions
as well as the Landau-Pomeranchuk-Migdal (LPM) effect~\cite{lpm} into account.
For the photonuclear cross-section, we
chose the extrapolation recommended by the Particle Data
Group ($\sigma^{\rm PDG}$)~\cite{pdg,cud}
shown in Fig.~\ref{fig-sigma_phn}.
The influence of assuming different extrapolations
is discussed below.
Hadron interactions were simulated with
QGSJET~01~\cite{qgsjet01}, and for energies $<$80~GeV with
GHEISHA~\cite{gheisha}.
\begin{figure}
\centerline{
\includegraphics[width=7.3cm]{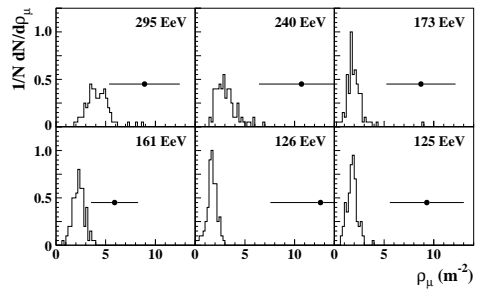}
}
\vskip -0.3 cm
\caption{
Observed muon densities (points with error bars) compared
to the muon densities expected for primary photons (histograms)
for the six events. Assigned to each event the primary
energy (see Tab.~\ref{table1}).
The measured muon densities are larger than predicted for primary
photons.
}
\label{fig-mudens}
\vskip -0.3 cm
\end{figure}
The distribution $\rho_{j}^{\rm s}$ of simulated muon densities
obtained from CORSIKA for each AGASA event is given
in Fig.~\ref{fig-mudens} together with the data.
The average values $<$$\rho_j^{\rm s}$$>$ and
standard deviations $\Delta \rho_j^{\rm s}$ are listed in Tab.~\ref{table1}.
The average muon densities for primary photons are
a factor 2-7 below the data.
To quantify the level of agreement between data and primary
photon expectation, a $\chi_j^2$ value is calculated
for each event $j$ as
\begin{equation}
\label{eq1}
\chi_j^2 = \frac{(\rho_j - <\rho_j^{\rm s}>)^2}
               { (\Delta \rho_j)^2 + (\Delta \rho_j^{\rm s})^2}
\end{equation}
with the measurement uncertainty
$\Delta \rho_j = 0.4 \cdot \rho_j$~\cite{agasa-data}.
To account for possible deviations of the simulated muon densities
from a Gaussian distribution,
the probability $p_j(\chi^2$$\ge$$\chi_j^2)$ of a photon-initiated
shower to yield a value $\chi^2 \ge \chi_j^2$ 
is determined by a Monte Carlo technique:
A simulated muon density is taken at random
from the distribution $\rho_{j}^{\rm s}$, a random shift is
performed according to the experimental resolution $\Delta \rho_j$,
and a $\chi^2$ value is calculated with Eq.~(\ref{eq1}), replacing
$\rho_j$ by the artificial muon density value.
Repeating this many times 
gives $p_j(\chi^2$$\ge$$\chi_j^2)$.
The values $\chi_j^2$ and $p_j$ are listed for the six events
in Tab.~\ref{table1}.
The probabilities $p_j$ range from 3\% to 21\%.

The combined probability $p(\chi^2$$\ge$$\sum_{j=1}^6 \chi_j^2)$
of six photon-initiated events
to yield a $\chi^2$ value larger or equal to the measured one is $p$ = 0.5\%.
Thus, it is unlikely
that all cosmic rays at these energies are photons (rejection with
99.5\% confidence), and it is possible to derive an upper limit
on the primary photon fraction $F_\gamma$.

It should be noted that, due to the small event statistics,
the upper limit cannot be smaller than a certain value.
Assuming a primary photon fraction $F_\gamma$,
a set of $n_{\rm m}$ primaries picked at random
{\it ab initio} does not contain any photon
with probability $(1-F_\gamma)^{n_{\rm m}}$.
For $n_{\rm m}$=6, this probability is $\simeq$5\% for $F_\gamma$=40\%.
Thus, in the present case only hypothetical photon
fractions $F_\gamma$$\ge$40\% could in principle
be tested at a confidence level of 95\%.

For deriving an upper limit $F_\gamma^{\rm ul}$$<$100\%,
scenarios have to be tested in which $n_\gamma$=$0\dots n_{\rm m}$
showers out of $n_{\rm m}$
events might be initiated by photons.
For a hypothetical photon fraction $F_\gamma$,
the probability $q$ that a set of $n_{\rm m}$
events contains $n_\gamma$ photons is
$q(F_\gamma,n_\gamma,n_{\rm m}) =
 F_\gamma^{n_\gamma} (1-F_\gamma)^{n_{\rm m} - n_\gamma}
(^{n_{\rm m}}_{n_\gamma})$.
This probability is multiplied by the probabilities
$p_\gamma(n_\gamma) \cdot p_{\overline \gamma}(n_{\rm m} - n_\gamma)$,
with $p_\gamma(n_\gamma)$ being the probability that the $n_\gamma$
most photon-like looking events are generated by photons,
and $p_{\overline \gamma}(n_{\rm m} - n_\gamma)$ being the probability
that the remaining $n_{\rm m}-n_\gamma$ events
are due to non-photon primaries.
$p_\gamma(n_\gamma)$ is determined by the MC technique as the
probability to obtain values 
$\chi^2 \ge \sum_{i=1}^{n_{\gamma}} \chi_{k_i}^2$,
with $p_\gamma(0)$=1 and
with $\chi_{k_i}^2$=$\chi_j^2$ from Tab.~\ref{table1},
where the index $k_1$ refers to the event with smallest value $\chi_j^2$,
and $\chi_{k_i}^2$$\le$$\chi_{k_{i+1}}^2$.
To derive an upper limit on photons, the probabilities
$p_{\overline \gamma}(n_{\rm m} - n_\gamma)$
are set to unity.
Summing over all possibilities $n_\gamma$=$0\dots n_{\rm m}$
then gives the probability $P(F_\gamma)$ to obtain
$\chi^2$ values at least as large as found in the data set,
\begin{equation}
P(F_\gamma) = \sum_{n_\gamma = 0}^{n_{\rm m}}
               q(F_\gamma,n_\gamma,n_{\rm m})
               \cdot p_\gamma(n_\gamma)
               \cdot p_{\overline \gamma}(n_{\rm m} - n_\gamma)~.
\end{equation}
This probability depends on the assumed photon fraction $F_\gamma$.
For the considered data set one obtains
$P(F_\gamma$$=$$51\%) = 10\%$ and $P(F_\gamma$$=$$67\%) = 5\%$.
Therefore, the upper limit on the primary photon fraction is 
$F_\gamma^{\rm ul} = 51\%$ (67\%) at 90\% (95\%) confidence level.

\begin{figure}
\centerline{
\includegraphics[width=7.5cm]{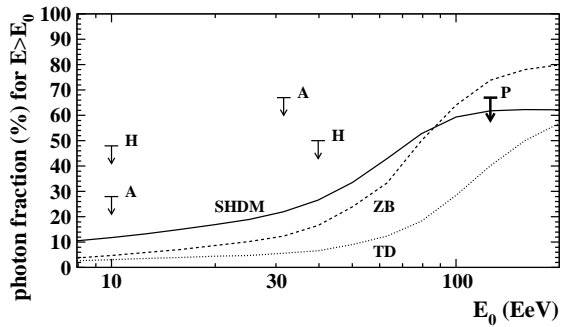}
}
\vskip -0.4cm
\caption{Upper limits (95\% CL) on cosmic-ray photon fraction
derived in the present analysis (P) and
previously from AGASA (A)~\cite{agasa-data}
and Haverah Park (H)~\cite{havpark} compared to some
predictions from super-heavy dark matter (SHDM)~\cite{berez04},
Z-burst (ZB) and topological defect (TD)~\cite{semsig} models.}
\label{fig-uplim}
\vskip -0.6cm
\end{figure}

The derived bound is the first experimental limit on the
photon contribution above the GZK cutoff energy.
The limit refers to the photon fraction integrated
above the primary photon energy that corresponds to the lowest-energy
event in the data sample, which in the present analysis is 125~EeV.
In Fig.~\ref{fig-uplim}, upper limits derived previously at lower
energy and the current bound are compared to some predictions based
on non-acceleration models. 
Models predicting photon dominance at highest energies are
disfavoured by the new upper limit.

The statistical stability of the upper limit can be tested by, e.g.,
omitting one event and calculating an upper limit with the
remaining five. Iterating through all six possibilities of rejecting
an event, the upper limits are between 61-78\% (95\% CL).
Alternatively, the 320~EeV Fly's Eye event
can be added to the event list with the photon probability
of 13\%~\cite{fe04}.
The upper limit then is 66\% (95\% CL). 
The result is quite stable, as the individual
photon probabilities do not differ much from each other.

The upper limit derived in the present analysis
is conservative with respect to
different sources of systematic uncertainties, since 
$p_j$ might be overestimated.
As an example, 
in the 295~EeV event data, muon detectors saturated and
the obtained $\rho_j$ might rather be regarded
as a lower limit~\cite{agasa-246,agasa-data}.
Concerning the simulations,
$\rho_j^{\rm s}$ is robust when changing the low-energy hadronic
interaction model~\cite{heckpylos}.
The applied high-energy model QGSJET~01 produces $\simeq$20-30\%
more muons~\cite{heckcern}
compared to SIBYLL~2.1~\cite{sibyll2.1} and also compared
to a preliminary version of QGSJET~II~\cite{qgsjet2}.
Smaller values of $\rho_j^{\rm s}$ or, in case of the 295~EeV event,
a larger value of $\rho_j$, would decrease $p_j$
and reduce the photon upper limit.

The derived upper limit is robust against reasonable variations of the
primary photon energy adopted. In general, a larger primary energy would
result in larger values of $\rho_j^{\rm s}$ and $p_j$.
We already accounted for a possible 20\% underestimation in case
of primary photons. It seems unlikely that an additional, systematic 
underestimation of reconstructed primary photon energies of
more than 20-30\% exists, also because of the stronger preshower effect
at increased energy that makes the profiles
of primary photon showers more similar to hadron-initiated events.
In turn, a rescaling of AGASA energies to smaller values would make
the muon densities predicted for photons even more discrepant to
the data.

A considerable uncertainty exists in extrapolating
the photonuclear cross-section.
A stronger (weaker) increase of the cross-section with energy
than adopted in this work leads to larger (smaller)
values of $\rho_j^{\rm s}$.
We repeated the calculations with different cross-section assumptions.
The upper limit of 67\% (95\% CL)
changes little for modest variations
of the extrapolation. Adopting, for instance, the parametrization denoted
$\sigma^{\rm mod}$ in Fig.~\ref{fig-sigma_phn}, the upper limit
becomes 75\% (95\% CL).
However, as an illustration, assuming the extreme extrapolation
labeled $\sigma^{\rm extr}$ (Fig.~\ref{fig-sigma_phn}),
the simulated $\rho_j^{\rm s}$ are increased on average by 
70-80\% with respect to the calculation using $\sigma^{\rm PDG}$.
In such a scenario, one would obtain
$P(\sigma^{\rm extr}, F_\gamma$=100\%)$ \simeq 15\%$, and
no upper limit could be set with high confidence.
Also the
previous limits from
Haverah Park and AGASA data~\cite{havpark,agasa-data} would increase
when assuming $\sigma^{\rm extr}$.

In summary, we introduced a new method for deriving an upper limit on
the cosmic-ray photon fraction from air shower observations.
Applied to the highest-energy AGASA events, an upper limit
of 67\% (95\% CL) is obtained for cosmic rays $>$125~EeV.
This photon bound imposes constraints on non-acceleration models of
cosmic-ray origin, with possible implications also on the description
of the dark matter or inflation scenarios in these models.
Within the next few years, a considerable increase of EHE event
statistics is expected
from the HiRes detector~\cite{hires} and the
Pierre Auger Observatory~\cite{auger}. Thus,
even more stringent conclusions on EHE photons are possible by
performing an analysis as presented here.
It will be studied elsewhere,
to what extend a scenario of dominant
EHE photons together with a large photonuclear cross-section
can be tested with shower data.

{\it Acknowledgments.}
We are grateful to K.~Eitel, M.~Je$\dot{\textnormal{z}}$abek,
M.~Kachelrie{\ss}, S.S.~Ostapchenko, S.~Sarkar, K.~Shinoza\-ki,
and G.~Sigl for helpful comments.
This work was partially supported by the Polish State Committee for
Scientific Research under grants No.~PBZ~KBN~054/P03/2001 and 
2P03B~11024 and in Germany by the DAAD under grant No.~PPP~323.
MR is supported by the Alexander von Humboldt-Stiftung.


\begin{thebibliography}{99}

\bibitem{vulcano}
J.~Linsley, Phys.~Rev.~Lett.~{\bf 10}, 146 (1963).

\bibitem{flyseye}
D.J.~Bird {\it et al.}, Phys.~Rev.~Lett.~{\bf 71}, 3401 (1993);
D.J.~Bird {\it et al.}, Astrophys.~J.~{\bf 441}, 144 (1995). 

\bibitem{agasa-213}
N.~Hayashida {\it et al.}, Phys.~Rev.~Lett.~{\bf 73}, 3491 (1994);
M.~Takeda {\it et al.}, Phys.~Rev.~Lett.~{\bf 81}, 1163 (1998).

\bibitem{havpark}
M.~Ave {\it et al.},
Phys.~Rev.~Lett.~{\bf 85}, 2244 (2000);
M.~Ave {\it et al.},
Phys.~Rev.~{\bf D65}, 063007 (2002).

\bibitem{agasa-data} 
K.~Shinozaki {\it et al.}, Astrophys.~J.~{\bf 571}, L117 (2002).

\bibitem{hires}
R.U.~Abbasi {\it et al.}, Phys.~Rev.~Lett. {\bf 92}, 151101 (2004). 

\bibitem{yakutsk}
V.P.~Egorova {\it et al.},
astro-ph/0408493 (2004).

\bibitem{reviews} 
M.~Nagano, A.A.~Watson, Rev.~Mod.~Phys. {\bf 72}, 689 (2000);
J.~Cronin, astro-ph/0402487 (2004);
{\it ``Ultimate energy particles in the Universe''}
(eds.~M.~Boratav, G.~Sigl), C.R.~Physique {\bf 5} (2004).

\bibitem{gzk}
K.~Greisen, Phys.~Rev.~Lett.~{\bf 16}, 748 (1966);
G.T.~Zatsepin, V.A.~Kuzmin, JETP Lett.~{\bf 4} 78 (1966).

\bibitem{stecker}
F.W.~Stecker, Phys.~Rev.~Lett.~{\bf 21}, 1016 (1968).

\bibitem{exotic}
C.T.~Hill, Nucl.~Phys.~{\bf B224}, 469 (1983);
M.B.~Hindmarsh, T.W.B.~Kibble, Rep.~Prog.~Phys.~{\bf 58}, 477 (1995);
V.~Berezinsky, M.~Kachelrie{\ss}, A.~Vilenkin,
Phys.~Rev.~Lett.~{\bf 79}, 4302 (1997);
M.~Birkel, S.~Sarkar, Astropart.~Phys.~{\bf 9}, 297 (1998);
V.A.~Kuzmin, V.A.~Rubakov,
Phys.~At.~Nucl.~{\bf 61}, 1028 (1998);
Z.~Fodor, S.D.~Katz, Phys.~Rev.~Lett.~{\bf 86}, 3224 (2001);
S.~Sarkar, R.~Toldra, Nucl.~Phys.~{\bf B621}, 495 (2002);
C.~Barbot, M.~Drees, Astropart.~Phys.~{\bf 20}, 5 (2003);
S.~Sarkar, Acta Phys.~Polon.~{\bf B35}, 351 (2004).

\bibitem{berez04}
R.~Aloisio, V.~Berezinsky, M.~Kachelrie{\ss},
Phys. Rev. {\bf D69}, 094023 (2004).

\bibitem{semsig}
G.~Sigl, hep-ph/0109202 (2001).


\bibitem{bhat-sigl}
P.~Bhattacharjee, G.~Sigl, Phys.~Rep.~{\bf 327}, 109 (2000).

\bibitem{constr_princ}
V.S.~Berezinsky, A.Yu.~Smirnov, Ap.~Sp.~Sci.~{\bf 32}, 461 (1975);
G.~Sigl, Science {\bf 291}, 73 (2001).

\bibitem{egret-diff}
P.~Sreekumar {\it et al.}, Astrophys.~J.~{\bf 494}, 523 (1998);
A.W.~Strong {\it et al.}, astro-ph/0405441 (2004).

\bibitem{sem-sigl} 
D.V.~Semikoz, G.~Sigl, JCAP {\bf 0404}, 003 (2004).

\bibitem{kachelr}
M.~Kachelrie{\ss}, C.R.~Physique {\bf 5}, 441 (2004).

\bibitem{agasa-excess}
M.~Takeda {\it et al.}, Astropart.~Phys.~{\bf 19}, 447 (2003);
N.~Hayashida {\it et al.}, astro-ph/0008102 (2000);
M.~Takeda {\it et al.}, Astrophys.~J.~{\bf 522}, 225 (1999);
http://www-akeno.icrr.u-tokyo.ac.jp/AGASA

\bibitem{agasa}
N.~Chiba {\it et al.}, Nucl.~Instr.~Meth.~{\bf A311}, 338 (1992);
H.~Ohoka {\it et al.}, Nucl.~Instr.~Meth.~{\bf A385}, 268 (1997).

\bibitem{latdistr}
N.~Hayashida {\it et al.}, J.~Phys.~{\bf G21}, 1001 (1995).

\bibitem{preshower}
P.~Homola {\it et al.},
astro-ph/0311442 (2003).

\bibitem{corsika}
D.~Heck {\it et al.},
Report {\bf FZKA 6019},
and D.~Heck, J.~Knapp, Report {\bf FZKA 6097},
Forschungszentrum Karls\-ruhe (1998).

\bibitem{egs}
W.R.~Nelson, H.~Hirayama, D.W.O.~Rogers, Report
{\bf SLAC 265}, Stanford Linear Accelerator Center (1985).

\bibitem{lpm} 
L.D.~Landau, I.Ya.~Pomeranchuk, Dokl.~Akad.~Nauk SSSR
{\bf 92}, 535 \& 735 (1953) (in Russian);
A.B.~Migdal, Phys.~Rev.~{\bf 103}, 1811 (1956).

\bibitem{pdg}
S.~Eidelmann {\it et al.}, Particle Data Group,
Phys. Lett. {\bf B592}, 1 (2004).

\bibitem{cud}
J.R.~Cudell {\it et al.}, Phys.~Rev.~{\bf D65}, 074024 (2002).

\bibitem{bezbug}
L.~Bezrukov, E.~Bugaev, Sov.~J.~Nucl.~Phys.~{\bf 33}, 635 (1981).

\bibitem{donland}
A.~Donnachie, P.~Landshoff, Phys.~Lett.~{\bf B518}, 63 (2001).

\bibitem{landboern}
H.~Genzel {\it et al.}, in: Landolt-B\"ornstein, New Series 
{\bf I/8}, Springer, Berlin (1973). 

\bibitem{qgsjet01}
N.N.~Kalmykov, S.S.~Ostapchenko, A.I.~Pavlov,
Nucl.~Phys.~B (Proc.~Suppl.) {\bf 52}, 17 (1997).

\bibitem{gheisha}
H.~Fesefeldt, Report  {\bf PITHA-85/02}, RWTH Aachen (1985).

\bibitem{fe04}
M.~Risse {\it et al.},
Astropart.~Phys.~21, 479 (2004).

\bibitem{agasa-246}
N.~Sakaki {\it et al.}, Proc.~27th Int.~Cosmic Ray Conf.,
Hamburg, 329 (2001).

\bibitem{heckpylos}
D.~Heck, astro-ph/0410735 (2004).

\bibitem{heckcern}
D.~Heck, M.~Risse, J.~Knapp,
Nucl.~Phys.~B (Proc.~Suppl.) {\bf 122}, 451 (2004).

\bibitem{sibyll2.1}
R.~Engel, T.K.~Gaisser, P.~Lipari, T.~Stanev, Proc.~26th
Int.~Cosmic Ray Conf., Salt Lake City, 415 (1999).

\bibitem{qgsjet2}
S.S.~Ostapchenko, hep-ph/0501093 (2005).

\bibitem{auger}
J.~Abraham {\it et al.}, Auger Collaboration,
Nucl.~Instrum.~Meth.~{\bf A523}, 50 (2004);
J.~Bl\"umer for the Auger Collaboration,
J.~Phys.~{\bf G29}, 867 (2003).

\end{thebibliography}
\end{document}